\documentstyle[twoside,fleqn,npb,epsfig]{article}
\newcommand{\AmS}{{\protect\the\textfont2
  A\kern-.1667em\lower.5ex\hbox{M}\kern-.125emS}}

\title{Combined analysis of diffractive and inclusive structure functions 
in the semiclassical framework}

\author{W.~Buchm\"uller\address{Deutsches Elektronen-Synchrotron DESY, 
Notkestr. 85, 22603 Hamburg, Germany}, 
T.~Gehrmann\address{Institut f\"ur Theoretische Teilchenphysik, 
Univ. Karlsruhe, Engesser Str. 7, 76131 Karlsruhe, Germany} and 
A.~Hebecker\address{Institut f\"ur Theoretische Physik der 
Univ.~Heidelberg, Philosophenweg 16, 69120 Heidelberg, Germany}}

\begin{document}

\begin{abstract}
Small-$x$ DIS is described as the scattering of a partonic fluctuation of 
the photon off a superposition of target color fields. Diffraction occurs 
if the emerging partonic state is in a color singlet. Introducing a specific 
model for the averaging over all relevant color field configurations, both 
diffractive and inclusive parton distributions at some low scale $Q_0^2$ 
can be calculated. A conventional DGLAP analysis results in a good 
description of diffractive and inclusive structure functions at higher 
values of $Q^2$. 
\end{abstract}

\maketitle

At this workshop, several numerical analyses of recent precise measurements 
of the diffractive structure function~\cite{h1,zeus} have been 
reported~\cite{sop,wue,pes,ing}. In this contribution, a combined 
description of inclusive and diffractive DIS in the semiclassical framework 
is discussed~\cite{bgh}. 

\begin{figure*}[t]
\begin{center}
\vspace*{-.5cm}
\parbox[b]{12cm}{\psfig{width=12cm,file=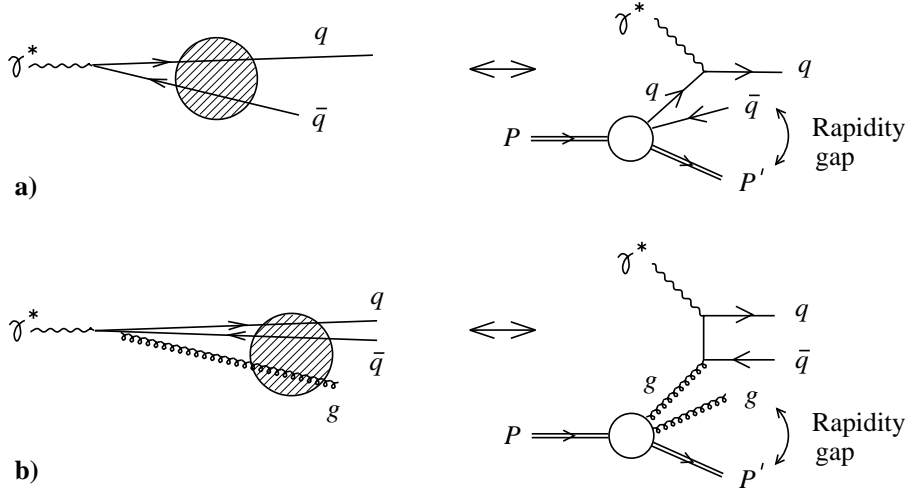}}\\
\vspace{-.8cm}
\end{center}
\caption{Diffractive DIS in the proton rest frame (left) and 
the Breit frame (right); asymmetric quark fluctuations correspond to 
diffractive quark scattering, asymmetric gluon fluctuations to diffractive 
boson-gluon fusion.}
\label{f2d}
\end{figure*}

From the target rest frame point of view, leading order diffractive DIS 
is the color singlet production of a $q\bar{q}$ pair, as shown on the l.h. 
side of Fig.~\ref{f2d}a. The process is dominated by kinematic 
configurations corresponding to Bjorken's aligned jet model, i.e., one of 
the quarks carries most of the photon's longitudinal momentum and the 
transverse momenta are small. The dependence of the cross section on the 
target color field is encoded in the expression 
\begin{equation}
\int d^2x_\perp \,\mbox{tr}W_{x_\perp}(y_\perp)\,\mbox{tr}
W_{x_\perp}^\dagger(y_\perp')\,,\label{wdef}
\end{equation}
where the function 
\begin{equation}
W_{x_\perp}(y_\perp)=U(x_\perp)U^{\dagger}(x_\perp+y_\perp)-1\label{um}
\end{equation}
is built from two SU(3) matrices, $U$ and $U^\dagger$, corresponding to 
the non-Abelian phase factors picked up by the quark and antiquark 
penetrating the color field at transverse positions $x_\perp$ and 
$x_\perp+y_\perp$. 

In the Breit frame, leading order diffractive DIS is most naturally 
described by photon-quark scattering, with the quark coming from the 
diffractive parton distribution of the target hadron~\cite{bs}. This is 
illustrated on the r.h. side of Fig.~\ref{f2d}a. Identifying the leading 
twist part of the $q\bar{q}$ pair production cross section (l.h. side of 
Fig.~\ref{f2d}a) with the result of the conventional partonic calculation 
(r.h. side of Fig.~\ref{f2d}a), the diffractive quark distribution of 
the target is expressed in terms of the color field dependent function 
given in Eq.~(\ref{wdef}). 

Similarly, the cross section for the color singlet production of a 
$q\bar{q}g$ state (l.h. side of Fig.~\ref{f2d}b) is identified with the 
boson-gluon fusion process based on the diffractive gluon distribution of 
the target (r.h. side of Fig.~\ref{f2d}b). This allows for the calculation 
of the diffractive gluon distribution in terms of a function similar to 
Eq.~(\ref{wdef}) but with the $U$ matrices in the adjoint representation. 

In the semiclassical approach, the cross sections for inclusive DIS are 
obtained from the same calculations as in the diffractive case where, 
however, the color singlet condition for the final state parton 
configuration is dropped. As a result, the $q\bar{q}$ production cross 
section (cf. the l.h. side of Fig.~\ref{f2d}a) receives contributions from 
both the aligned jet and the high-$p_\perp$ region. In the latter, the 
logarithmic $d p_\perp^2/p_\perp^2$ integration gives rise to a $\ln Q^2$ 
term in the full cross section. 

In the leading order partonic analysis, the full cross section is described 
by photon-quark scattering. The gluon distribution is responsible for the 
scaling violations at small $x$, $\partial F_2(x,Q^2)/\partial\ln Q^2 
\sim xg(x,Q^2)$. Thus, the semiclassical result for $q\bar{q}$ production, 
with its $\ln Q^2$ contribution, is sufficient to calculate both the 
inclusive quark and the inclusive gluon distribution. The results are again 
expressed in terms of the function in Eq.~(\ref{wdef}) where now the color 
trace is taken {\it after} the two $W$ matrices (corresponding to the 
amplitude and its complex conjugate) have been multiplied. 

To obtain explicit formulae for the above parton distributions, a model for 
the averaging over the color fields, which underlie the eikonal factors in 
Eq.~(\ref{um}), has to be introduced. In the case of a very large 
hadronic target~\cite{mv}, such a model is naturally obtained from the 
observation that, even in the aligned jet region, the transverse separation 
of the $q\bar{q}$ pair remains small~\cite{hw}. This is a result of the 
saturation of the dipole cross section at smaller dipole size. Under the 
additional assumption that color fields in distant regions of the large 
target are uncorrelated, a simple Glauber-type exponentiation of the 
averaged local field strength results in explicit formulae for all the 
relevant functions of the type shown in Eq.~(\ref{wdef}). 

Thus, diffractive and inclusive quark and gluon distributions at some 
small scale $Q_0^2$ are expressed in terms of only two parameters, the 
average color field strength and the total size of the large target 
hadron. The energy dependence arising from the large-momentum cutoff 
applied in the process of color field averaging can not be calculated from 
first principles. It is described by a $\ln^2 x$ ansatz, consistent with 
unitarity, which is universal for both the inclusive and diffractive 
structure function~\cite{b}. This introduces a further parameter, the 
unknown constant that comes with the logarithm. 

\begin{figure}[t]
\vspace*{-.35cm}
\begin{center}
\parbox[b]{5.5cm}{\psfig{file=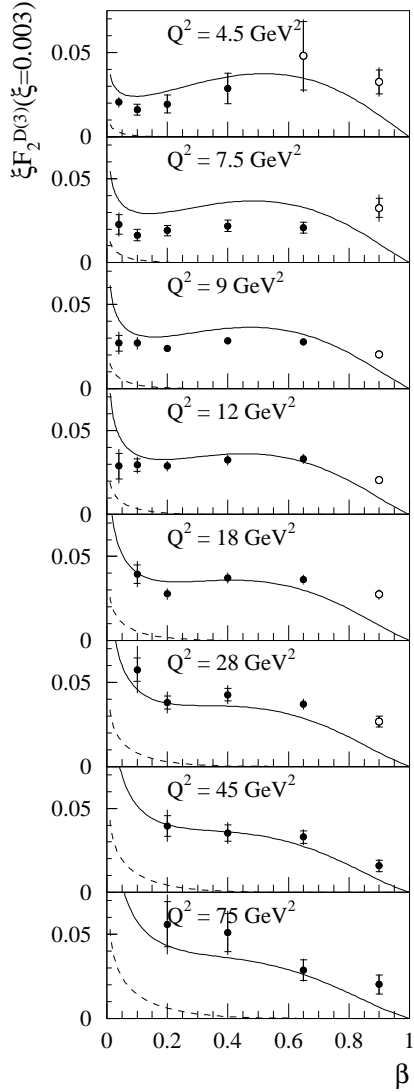,width=5.5cm}}\\
\vspace{-1cm}
\end{center}
\caption{The diffractive structure function $F_2^{D(3)}(\xi,\beta,Q^2)$ with 
data from H1~\cite{h1}. Open circles correspond to $M^2\leq 4$~GeV$^2$. 
The charm content is indicated as a dashed line.}
\label{h1}
\vspace*{-.2cm}
\end{figure}

\begin{figure}[t]
\vspace*{-3.7cm}
\begin{center}
\parbox[b]{5.5cm}{\psfig{file=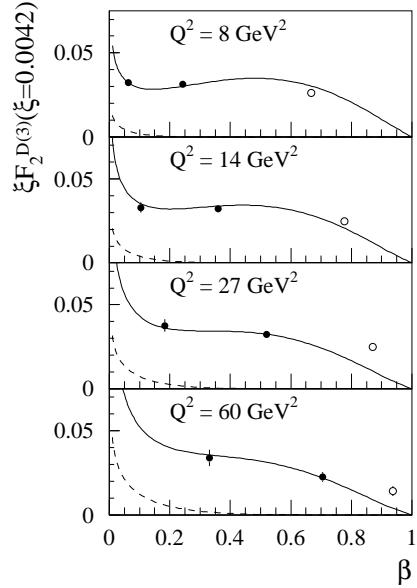,width=5.5cm}}\\
\vspace{-4.5cm}
\end{center}
\caption{Diffractive structure function $F_2^{D(3)}$ (conventions of 
Fig.~\ref{h1}) with data from ZEUS~\cite{zeus}.}
\label{zeus}
\vspace*{-.2cm}
\end{figure}

A conventional leading order DGLAP analysis of data at small $x$ and 
$Q^2>Q_0^2$ results in a good four parameter fit ($Q_0$ being the fourth 
parameter) to both the inclusive and diffractive structure function. 
Diffractive data with $M^2<4\,\mbox{GeV}^2$ 
is excluded from the fit since higher twist effects are expected to affect 
this region. As an illustration, the $\beta$ dependence of $F_2^{D(3)}$ at 
different values of $Q^2$ is shown in Figs.~\ref{h1} and \ref{zeus} 
(see~\cite{bgh} for further plots, in particular of the inclusive structure 
function, and more details of the analysis). 

Finally, two important qualitative features of the approach should be 
emphasized. First, the diffractive gluon distribution is much larger than 
the diffractive quark distribution, a result reflected in the pattern of 
scaling violations of $F_2^{D(3)}$. This feature is also present in the 
analysis of~\cite{hks}, where, in contrast to the present approach, the 
target is modelled as a small color dipole. Second, the inclusive gluon 
distribution, calculated from $q\bar{q}$ pair production at high $p_\perp$ 
and determined by the small-distance structure of the color field, is large 
and leads to the dominance of inclusive over diffractive DIS.

\end{document}